# Motion Correction of 3D Dynamic Contrast-Enhanced Ultrasound Imaging without Anatomical Bmode Images

Jia-Shu Chen*, Maged Goubran* Ph.D., Gaeun Kim, Jürgen K. Willmann M.D.,
Michael Zeineh M.D., Ph.D., Dimitre Hristov Ph.D., Ahmed El Kaffas Ph.D.
*These authors contributed equally.

*Abstract*— Dynamic contrast-enhanced ultrasound (DCE-US) is highly susceptible to motion artifacts arising from patient movement, respiration, and operator handling and experience. Motion artifacts can be especially problematic in the context of perfusion quantification. In conventional 2D DCE-US, motion correction algorithms take advantage of accompanying side-by-side anatomical Bmode images that contain time-stable features. However, current commercial models of 3D DCE-US do not provide side-by-side anatomical Bmode images, which makes motion correction challenging. This work introduces a novel motion correction (MC) algorithm for 3D DCE-US and assesses its efficacy when handling clinical data sets. In brief, the algorithm uses a pyramidal approach whereby short temporal windows consisting of 3-6 consecutive frames are created to perform local registrations, which are then registered to a master reference derived from a weighted average of all frames. We evaluated the algorithm in 8 patients with metastatic lesions in the liver using the Philips X6-1 matrix transducer at a frame rate of 1-3 Hz. We assessed improvements in original vs. motion corrected 3D DCE-US cine using: i.) frame-to-frame volumetric overlap of segmented lesions, ii.) normalized correlation coefficient (NCC) between frames (similarity analysis), and iii.) sum of squared errors (SSE), root-mean-squared error (RMSE), and r-squared ($R^2$) quality-of-fit from fitted time-intensity curves (TIC) extracted from a segmented lesion. Overall, results demonstrate significant decreases in 3D DCE-US motion after applying the proposed algorithm. We noted significant improvements in frame-to-frame lesion overlap across all patients, from 68%±13% without correction to 83%±3% with motion correction ($p = 0.023$). Frame-to-frame similarity as assessed by NCC also significantly improved on two different sets of time points from 0.694±0.057 (original cine) to 0.862±0.049 (corresponding MC cine) and 0.723 ±0.066 to 0.886 ±0.036 ($p \le 0.001$ for both). TIC analysis displayed a significant decrease in RMSE ($p = 0.018$) and a significant increase in $R^2$ goodness-of-fit ($p = 0.029$) for the patient cohort.

## I. INTRODUCTION

The introduction of contrast-enhanced ultrasound imaging has significantly expanded the diagnostic potential of ultrasound as a non-invasive and inexpensive imaging modality (1). Imaging tissue perfusion is feasible using dynamic contrast-enhanced ultrasound (DCE-US) methods in contrast-mode to isolate contrast signal from tissue signals in several abdominal indications, as well as other anatomical sites accessible to ultrasound, and has been adopted in global clinics for over a decade now (2). While both CT and MRI imaging modalities also offer contrast-enhanced tissue perfusion measurements, ultrasound-based contrast imaging is advantageous clinically because it is inexpensive relative to other imaging modalities, without radiation, available at the bedside, and offers a uniquely intravascular contrast agent that does not leak out of tumor permeable vessels. More recently, three-dimensional (3D) DCE-US was introduced with the availability of contrast-mode imaging on commercially-available clinical matrix transducers. 3D DCE-US offers volumetric imaging that minimizes sampling errors and operator-influence on quantification; this is particularly important in longitudinal imaging applications such as cancer treatment monitoring (3). In contrast, imaging tumors via conventional 2D DCE-US is prone to sampling errors and can produce biased quantitative results because of plane-to-plane perfusion variation and tumor heterogeneity (4). 3D DCE-US imaging mitigates these errors by imaging the tumor as a whole through dynamic volumetric images (5).

Several quantification techniques have been introduced to characterize tissue perfusion properties from DCE-US imaging studies (1-6). Perfusion changes, which can be quantified longitudinally using CE-US methods, are indicative of both disease type and treatment response (7-10). However, the quality of quantification is affected by several imaging and clinical limitations. In particular, ultrasound-based dynamic imaging is highly susceptible to motion artifacts arising from patient movement, respiration, and peristalsis, as well as operator hand stability (11). Motion artifacts can be especially hampering in lengthy acquisitions and in the context of quantification by introducing significant errors to the quantified contrast signal. In

- Jia-Shu Chen is with the Department of Neuroscience at Brown University, Providence, RI USA (e-mail: jia-shu_chen@brown.edu).
- Maged Goubran is with the Department of Radiology at Stanford University, Stanford, CA USA (email: mgoubran@stanford.edu).
- Gaeun Kim is with the Department of Radiology at Stanford University, Stanford, CA USA (e-mail: gaeunkim@stanford.edu).
- Ahmed El Kaffas is with the Department of Radiology, Molecular Imaging Program at Stanford School of Medicine, Stanford University, Stanford, CA USA (email: elkaffas@stanford.edu).

- Dimitre Hristov is with the Department of Radiation Oncology – Radiation Physics at Stanford School of Medicine, Stanford University, Stanford, CA USA (email: dimitre.hristov@stanford.edu).
- Michael Zeineh is with the Department of Radiology at Stanford University, Stanford, CA USA (email: mzeineh@stanford.edu).
- Jürgen K. Willmann was with the Department of Radiology, Molecular Imaging Program at Stanford School of Medicine, Stanford University, Stanford, CA USA a study investigating the ability of conventional 2D DCE-US imaging to measure tumor response to targeted therapy in renal cell carcinoma patients, the investigators observed significant differences due to respiratory motion in perfusion parameters after bolus time-intensity analysis (6). Another study investigating ultrasound imaging of focal liver lesions encountered technical failures due to motion artifacts from patient movement (12). These studies have concluded that motion correction (MC) techniques that account for motion artifacts are essential for reliable usage of DCE-US imaging in cancer treatment monitoring (13).

In conventional DCE-US, motion artifacts can originate from two sources - the respiratory cycle of patients being examined, and/or probe motion by the operator or the patient moving - the latter is especially challenging in longer scan sequences such as is often the case with contrast imaging where the probe can also drift out of the original target plane. In our use-case scenario, contrast-cine in 3D last 3-10 min (bolus or infusion) [REF]. A common approach to mitigate motion is to ask the patient to breath-hold, however, this only reduces respiratory motion in roughly the first 5-20 seconds of a cine sequence, and depends largely on a patients' ability to hold breath; breath hold can also introduce large motion artifacts when the patient resumes breathing with a large breath. Several gating techniques have also been reported to focus quantification only on frames within the same respiratory cycle [REF]; however, gating is non-ideal in quantification approaches as it reduces the temporal resolution of contrast kinetics. This is especially challenging with current clinical implementation of 3D DCE-US where the frame rate is already below 3 Hz. Gating can also be time consuming if done manually, and does not eliminate non-cyclic motion (i.e. operator or patient direct movements).

Commercial software solutions such as Qvue or [BRACCO software] offer a suite of quantification tools that include motion correction; these tend to rely on side-by-side Bmode images available in 2D DCE-US, and lock in on stable image features in the Bmode data to register all frames back to a reference frame [REF]. However, it is important to note that none of the clinically available commercial implementations of 3D DCE-US currently offer a side-by-side Bmode. Several novel motion correction algorithms for DCE-US have been developed that do not rely on Bmode data, and instead require modelling of the contrast flow from the TIC, and noise signal within an ROI [REF:Barros papers]. Nonetheless, while incredibly promising, these are mainly designed with 2D imaging in mind, and are heavily limited by out-of-plane motion, despite attempts to minimize these during acquisition or through gating, especially in non-cyclic conditions [REF: 13-14]. In 3D, one can utilize adapted registration methods and enable the use of more flexible transform space to improve motion correction, all while designing algorithms for the current limitations of 3D DCE-US implementations that include: 1) lack of side-by-side Bmode, and 2) low frame rate.

An inability to implement out-of-plane MC is not an issue with 3D DCE-US studies, however, no such side-by-side Bmode images are available in current commercial implementations of 3D DCE-US imaging. This lack of anatomical accompanying Bmode images pose a challenge for using stable anatomical features as a reference for registration and motion compensation. Additionally, the relatively slow frame rates (compared to 2D imaging) (4) and highly dynamic nature of microbubble-contrast agents as they wash-in and -out makes it very challenging to register subsequent frames to one another based off of proximity in time. The need for MC in DCE-US is especially important in quantitative clinical applications where perfusion parameters are used to measure relative changes in blood flow (i.e. treatment monitoring) (15) and more so when parametric maps of perfusion parameters are being generated on a voxel-by-voxel basis. The importance of MC in a voxel-by-voxel quantification setting was observed when we analyzed our clinical data and noticed that single voxel noise artifacts significantly affects voxel-wise parametric map generation and decreases time-intensity curve quality.

In this paper, we detail the development of a novel MC algorithm for 3D DCE-US data unaccompanied by Bmode anatomical imaging that improves the reliability of bolus time-intensity analysis for perfusion parameters. We first describe the details of the MC algorithm, and subsequently demonstrate its efficacy using 8 clinical data sets of liver metastasis 3D DCE-US imaging studies. To the best of our knowledge, no previous investigators have developed a MC algorithm for 3D DCE-US imaging of bolus microbubble injections.

## II. MATERIALS AND METHODS

### A. Clinical Case Analysis

The MC algorithm was evaluated using eight human 3D-DCE US imaging studies of liver metastasis. All studies were approved by the Stanford University IRB. The imaging studies were acquired from patients who provided written consent, were 18 years of age or older, and presented with at least one liver metastasis that was within the range of 1-14 cm in diameter and confirmed using MRI/CT imaging studies. The 3D-DCE US scanning was performed by an experienced sonographer using the commercial Philips X6-1 MHz xMATRIX array transducer at a setting of 1-3 Hz. Each original imaging study (pre-MC) was processed using the algorithm to produce a motion corrected imaging study (post-MC).

### B. Positional Evaluation of MC Improvement

In imaging studies affected by movement, the lesion often changes position within the 3D space captured by the transducer as a result of either patient or operator movement. In order to quantify the positional shift of the lesion from frame to frame, the lesions (tumors) were manually segmented using ITK-SNAP (18). Using MeVisLab (MeVis Medical Solutions, Version 3.0) quantification modules, the 3D lesion volume overlap was compared between frames within the sequence in order to determine the consistency of the lesion's position pre-MC and post-MC.

### C. Similarity Evaluation of MC Improvement

To determine whether MC resulted in a significant improvement that persisted throughout the entire imaging study, image similarity between neighboring frames was quantified within the imaging studies pre-MC and post-MC using normalized cross-correlation.

### D. Bolus Time-Intensity Evaluation of MC Improvement

The presence of motion artifacts impairs the quality and reproducibility of quantitative data when generating time-intensity curves (TIC) for 3D

DCE-US imaging studies. When motion artifacts are present, movement of the lesion of interest results in incorporation of noisy and irrelevant voxel intensity data into the TIC quantification software (19). Consequently, TIC data on tumor perfusion typically provides an inaccurate assessment of tumor response to cancer treatment. In order to evaluate improvement of TIC reliability, TICs were generated from the imaging studies pre-MC and post-MC using an in-house quantification module written in MeVisLab (MeVis Medical Solutions, Version 3.0) and MATLAB (Math Inc., Version 9.4). The quality and reliability of the TIC was assessed by measuring the sum of squared errors (SSE), the root-mean-squared-errors (RMSE), and $R^2$ values and comparing the values measured pre-MC and post-MC.

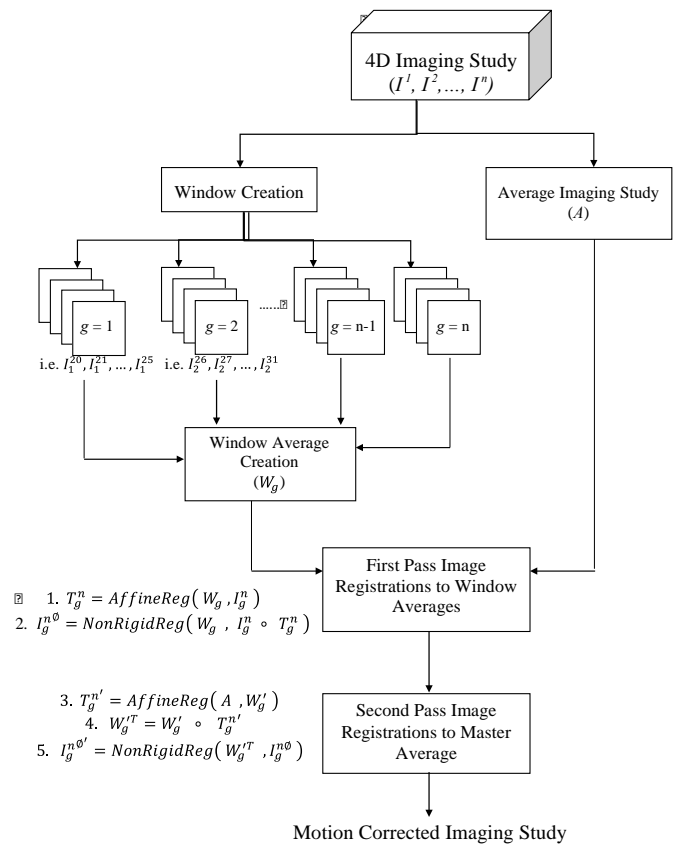

Fig. 1. Outline of the MC Algorithm

## III. MC ALGORITHM DESCRIPTION

The MC algorithm is based off of a two-pass window approach that relies on region-of-interest image masking, employs both affine and non-rigid image registration, and avoids the issue of variability between neighboring images biasing image registrations. Fig. 1 represents the schematic of the algorithm, and Table I. defines the symbols used in the description of the algorithm. We utilized the NiftyReg open-source medical image registration software (16) and the FSL FMRIB Software Library for Imaging Analysis (17) in our implementation and executed the algorithm on the Stanford Sherlock Computing Cluster [Add reference].

Each image acquisition represents a 4D cine consisting of 3D volumetric US image frames. In each imaging study, there are $N$ image frames, $I$, comprising the video, $I^{n \in \{1,...,N\}}$. All $n$ images are used to create a master reference image for later use, $A$, that is the average of all the 3D image frames in the acquisition. The algorithm corrects images following microbubble injection. Since microbubble injection occurs at different frames for each acquisition, the algorithm determines the starting image frame, $I^s$, for MC by searching for the first image with a mean voxel intensity that exceeds XX% of the mean voxel intensity in baseline images; we manually verfied that the correct frame was selected for each data set. The images identified for MC, $n \in \{1,...,N\text{-}s\}$, are then divided into $w$ short temporal windows of equal sizes, $W_{g \in \{1,...,G\}}$. Once the images have been categorized into specific windows, $I_w^n$, the images can then proceed to reference creation and image registration. The MC algorithm takes advantage of a series of affine and non-rigid image registrations between reference images and floating images (image that is restructured with respect to the reference image) in order to spatially realign and eliminate shape distortions within the images.

*1. First Pass:*

The first pass of image registration is intended to spatially realign and reshape the images with respect to the neighboring images within their respective windows. The initial image should ideally undergo a series of translations, rotations, resizing, shearing, and deformations that make it similar structurally and positionally to its neighboring images. Therefore, a window reference image, $W_g$, is created by averaging all the image frames comprising the window. The first pass of registrations begins with an affine registration of the images, $I_g^n$, to their window average, $W_g$, in order to create a spatially realigned image, $I_g^{n\prime}$. The image frame, $I_g^{n\prime}$, then undergoes a non-rigid registration using the affine transformation, $T_g^n$, as an initialization to create the final image product of the first pass, $I_g^{n\emptyset}$.

$$T_g^n = AffineReg\left( W_g \ , \ I_g^n \right) \qquad (1.)$$

$$I_g^{n\emptyset} = NonRigidReg\left( W_g \ , \ I_g^n \circ T_g^n \right) \quad (2.)$$

*2. Second Pass:*

The second pass's purpose is to realign and reshape the image product of the first pass with respect to the entire imaging study. Since the first pass created a new series of images, a new window reference image, $W_g{}'$, is created by averaging all the newly transformed images within the window. The new window averages are then registered using an affine transform to the original master average, $A$, to create window averages, $W_g{}'^T$, that are all spatially in sync with the average of the entire study. The second pass concludes with a non-rigid registration of the first pass image products, $I_g^{n\emptyset}$, to their final window averages, $W_g{}'^T$. The products of the second pass, $I_g^{n\emptyset\prime}$, should now be structurally and positionally similar to both images inside and outside their respective windows, thereby creating a motion corrected image study that occupies the same 3D space throughout the entire sequence.

$$T_g^{n\prime} = AffineReg\left( A \ , W_g{}' \right) \qquad (3.)$$

$$W_g{}'^T = W_g{}' \circ T_g^{n\prime} \qquad (4.)$$

$$I_g^{n\emptyset\prime} = NonRigidReg\left( W_g{}'^T \ , I_g^{n\emptyset} \right) \qquad (5.)$$

TABLE I
DESCRIPTORS FOR MC ALGORITHM

| Symbol | Meaning |
| --- | --- |
| $I$ | Image Frame |
| $n$ | Frame's time point in imaging acquisition |
| $N$ | Number of frames in imaging acquisition |
| $A$ | Imaging acquisition average |
| $s$ | Starting frame for MC |
| $g$ | Frame's window group |
| $G$ | Number of window groups |
| $W$ | Window Average |

| $T$ | Affine transform |
| $\emptyset$ | Non-Rigid deformation field |

### A.  MC Algorithm Technical Details

Image registrations performed in the algorithm were executed by the open-source NiftyReg medical image registration software (16). Briefly, the rigid registration relied on a symmetric block-matching algorithm and least-trimmed square regression. We employed a 12 degrees of freedom (affine) transform, with a normalized mutual information (NMI) similarity metric and a threshold of 10% outlier blocks in the optimization scheme. For the non-rigid (deformable) b-spline registration, the implementation performed a three-stage approach each with a higher resolution (grid), utilizing NMI similarity with 64 bins, a spline spacing of 5 voxels at the full resolution. For regularization, we used a bending energy penalty of 0.3 to constrain non-realistic deformations, and log of the Jacobian determinant penalty of 0.1. These parameters were optimized using a grid-search on 3 datasets.

To improve the quality of image registration, region-of-interest masks were created using the FSL FMRIB software library (17) and MeVisLab visualization (MeVis Medical Solutions, Version 3.0) to constrain the registration algorithms to the relevant image features. One mask was constructed per 4D acquisition. The algorithm pipeline was designed to be executed by the Stanford Sherlock Cluster, which provides 64000 MB of memory per node and 16 cores per node. The pipeline was rewritten to accommodate parallel processing within the Stanford Sherlock Cluster to decrease lengthy runtimes. Parallelization of the MC algorithm was able to reduce the processing times approximately four-fold on average (720 minutes to 180 minutes).

### IV.  Results

### A.  Positional Improvement Evaluation

Following MC, the volume overlap of the lesions increased significantly ($p = 0.023$), and the overall lesion remained in the same 3D position (as shown in **Fig. 2 and 3**). **Table 2** expresses the average percentage of lesion volume that overlaps with the other images in the sample sequence for all eight patient studies. For images that exhibited poor positional similarity and low volume overlap initially (~40-60%), MC was capable of significantly improving the positional similarity and volume overlap to exceed 80% for most data sets. For images that possessed high-volume overlap prior to MC, the images post-MC either improved slightly or remained at relatively the same volume overlap, indicating that the MC algorithm does not cause regression in images not heavily affected by motion artifacts. **Fig. 4** shows the improvement in lesion repositioning following MC.

### B.  Image Similarity Improvement Evaluation

The similarity analysis shows that there is an improvement in normalized correlation coefficients post-MC. **Table 2** shows that motion corrected images have higher image similarity, meaning that consecutive images were able to represent similarly positioned anatomy with respect to time following MC. Significant improvement ($p \leq 0.001$) in image similarity at two different sets of image frames (35-39 and 55-59) within the same 4D sequence persists throughout the imaging acquisition and not just at specific windows of the 4D sequence as seen in **Fig. 5**.

### C.  Time-Intensity Analysis Improvement Evaluation

MC reduces the average SSE and RMSE for all the generated TICs, as seen in **Table 3**. Furthermore, the $R^2$ value also increased significantly following MC, indicating that the fitted curve for TICs experienced less noisy and outlier data. The fitted curves for motion corrected data experienced less data spikes qualitatively and a decreased range of intensity values, as seen in **Fig. 6**. A lognormal model was used to fit our curves, as has been suggested to be an ideal curve fitting model for liver masses [REF].

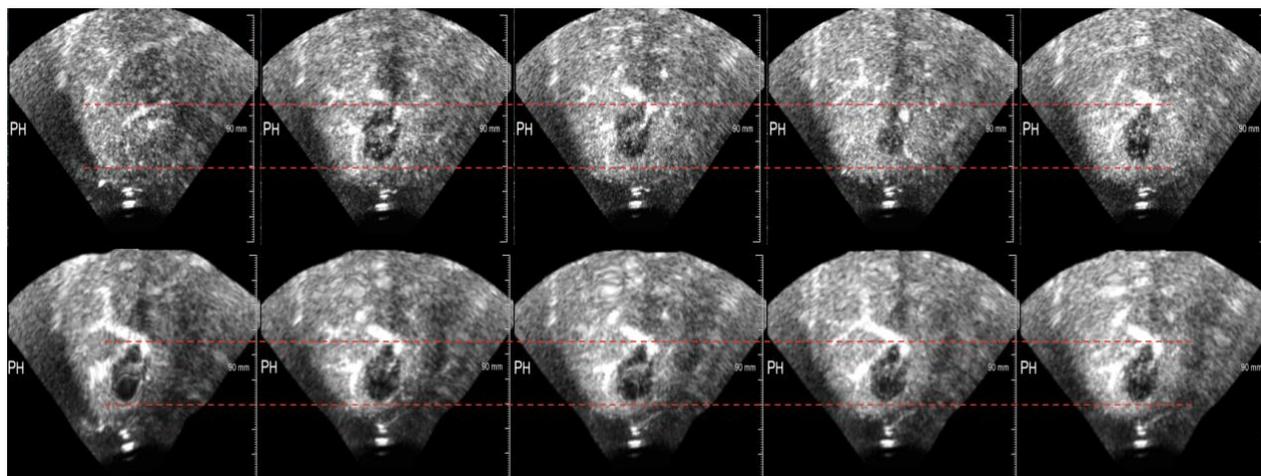

**Figure 2.** Frames from the Pre-MC imaging scan of P-01 (top) compared to their corresponding frames in the Post-MC imaging scan (bottom).

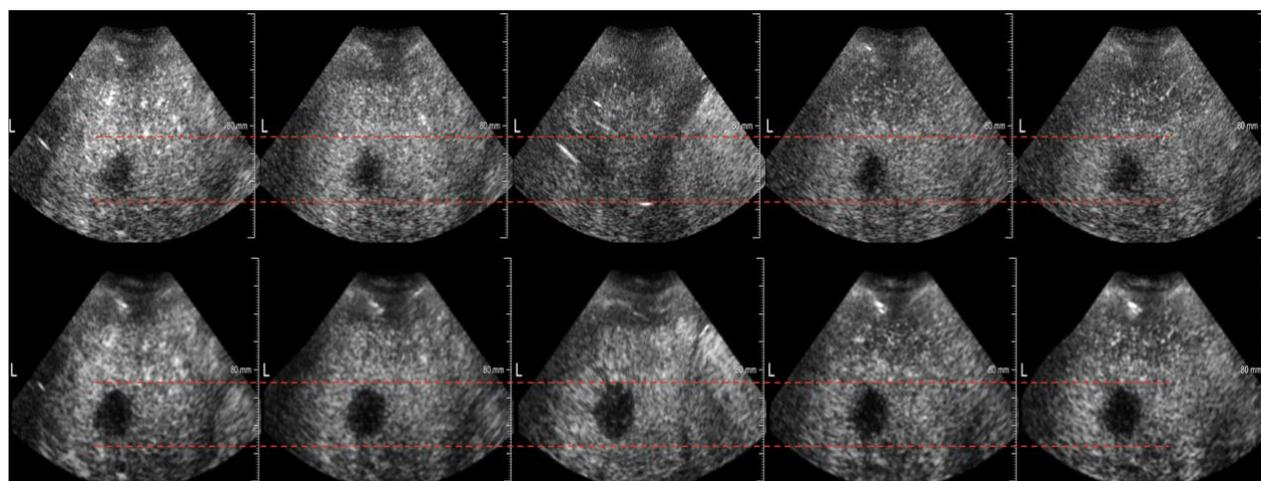

**Figure 3.** Frames from the Pre-MC imaging scan of P-04 (top) compared to their corresponding frames in the Post-MC imaging scan (bottom).

**TABLE 2. Results of Positional Improvement and Image Similarity Evaluation**

|  | Average Volume Overlap of Liver Lesion | | Average Image Similarity by Normalized Cross-Correlation (Frames 35-39) | | Average Image Similarity by Normalized Cross-Correlation (Frames 55-59) | |
|---|---|---|---|---|---|---|
|  | Pre-MC | Post-MC | Pre-MC | Post-MC | Pre-MC | Post-MC |
| Average | 68.39% ± 13% | 83.15% ± 3% | 0.723 ± 0.066 | 0.886 ± 0.036 | 0.694 ± 0.057 | 0.862 ± 0.049 |
| *p*-value | 0.023* | | 0.0001*** | | 4.30 x $10^{-5}$ *** | |

* $p \leq 0.05$, ** $p \leq 0.01$, *** $p \leq 0.001$

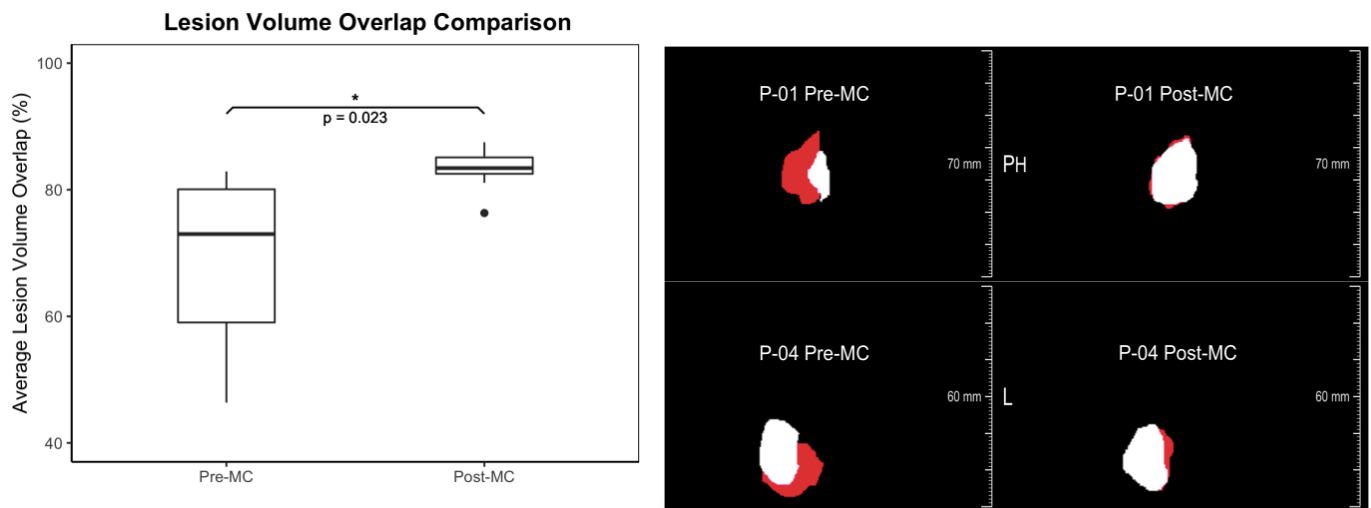

**Figure 4**. Box and whisker plot comparing Pre-MC and Post-MC average lesion volume overlap (left). Side-by-side comparison of two segmented lesions from neighboring frames of two different imaging studies overlaid on one another (right). The two top panels represent segmented lesions from P-01, while the two bottom panels represent segmented lesions from P-04. The left two panels are Pre-MC, while the right two panels are Post-MC. The white segmented lesion is overlapped on top of the red segmented lesion, therefore greater visualization of the red segmented lesion implies greater lesion displacement and poor concordance in tumor morphology between the two frames being compared.

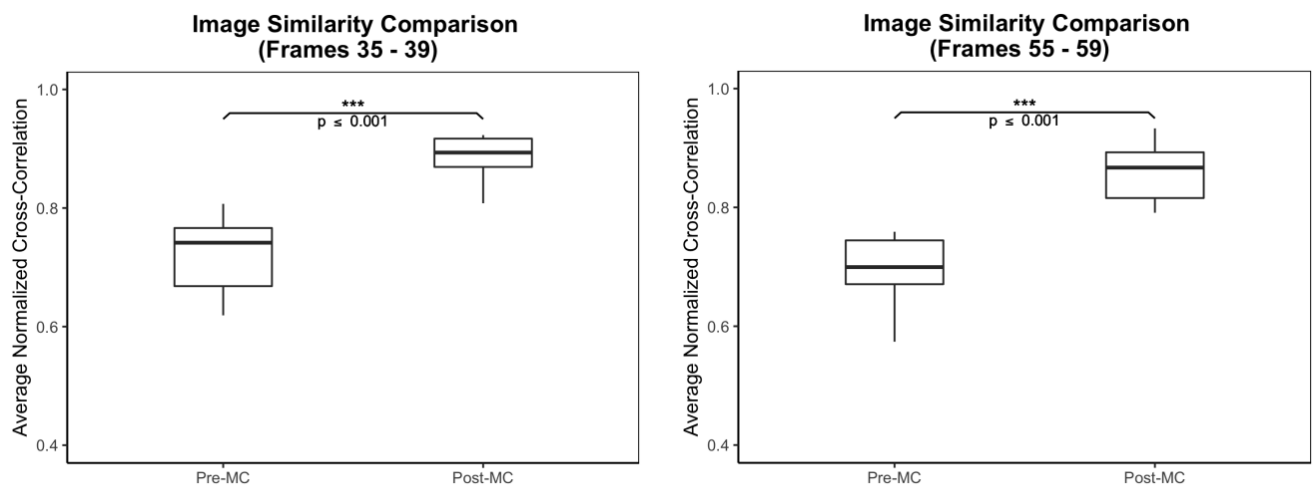

**Figure 5.** Box and whisker plots comparing the average normalized cross-correlation values for the Pre-MC and Post-MC imaging studies. The left plot represents the average normalized cross-correlation values for time frames 35-39, while the right plot represents the average normalized cross-correlation values for time frames 55-59 for each of the 8 patient imaging studies.

## TABLE 3. Results of Bolus Time-Intensity Metrics Evaluation

| | Sum of Squared Errors (SSE) | | Root Mean Square Error (RMSE) | | Coefficient of Determination ($R^2$) | |
|---|---|---|---|---|---|---|
| | Pre-MC | Post-MC | Pre-MC | Post-MC | Pre-MC | Post-MC |
| Average | $0.472 \pm 0.394$ | $0.166 \pm 0.073$ | $0.078 \pm 0.025$ | $0.049 \pm 0.012$ | $0.861 \pm 0.062$ | $0.932 \pm 0.044$ |
| *p*-value | 0.065 | | 0.018* | | 0.029* | |

\* $p \leq 0.05$, \*\* $p \leq 0.01$, \*\*\* $p \leq 0.001$

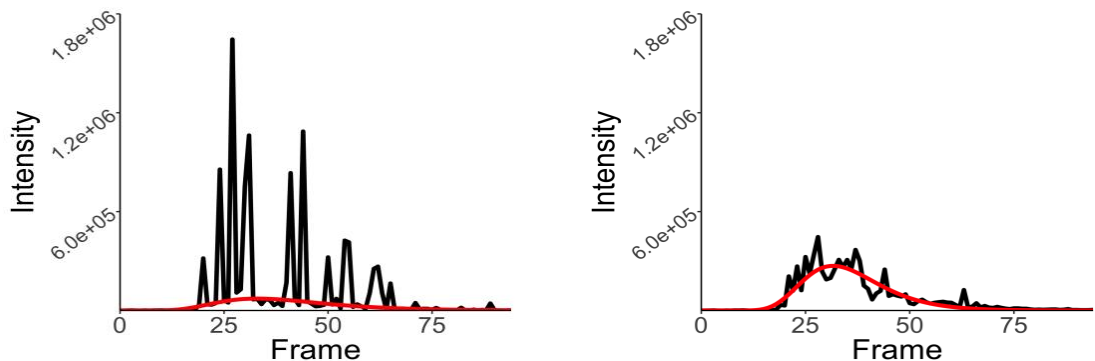

**Figure 6.** Comparison of fitted time-intensity curves at the tumor region of interest for P01 prior to MC (left) and after MC (right).

## V. Discussion

DCE-US imaging is highly susceptible to changes in the imaging window and transducer beam stemming from operator and/or subject movement (11). In the context of quantitative analysis of volumetric (3D) imaging studies for cancer treatment monitoring, minor changes in the orientation of the transducer can be problematic for the reliability and reproducibility of the perfusion parameters extracted. Studies into quantitative mapping of tumor vascularity have reported errors ranging from 6.4% to 40.3% due to millimeter-sized deviations in transducer positioning (20). As a result, we developed a MC algorithm that aims to mitigate the degree of misalignment found inherently in each pre-processed imaging study.

For the first evaluation methodology, presented in **Table 2** and **Figure 4**, volume overlap for a window series of five volumetric imaging frames per patient increased significantly following MC. Additionally, tumor morphology that previously differed qualitatively also improved in similarity with the lesions closely resembling its neighboring frames. This is advantageous during voxel-by-voxel quantification of DCE-US imaging studies for perfusion parameters because it reduces the potential for non-lesion intensity signals to interfere with the region-of-interest and cause the quantification software to misinterpret surrounding noise as relevant perfusion. This is highlighted in one case investigating the quantification of tumor microvasculature where more than 50% of acquisitions were not quantifiable due to excess noise caused by motion that made the lesions non-stationary and the TICs highly sporadic (21).

Image similarity comparisons using normalized cross correlation were employed to confirm that voxel intensities in one frame corresponded to those of the neighboring frames.

As seen in **Table 2** and **Figure 5**, the magnitude of the average image similarity per patient improved significantly for two different window series, implying that the voxel correlation between subsequent frames in the imaging acquisition has improved. With improved voxel alignment from frame to frame within a volumetric imaging study, an improvement in voxel-by-voxel perfusion quantification should be expected. Due to improved concordance and alignment between voxels going from frame to frame, the maximum intensity values in time-intensity curves should decrease because the interference of misaligned, non-lesion voxel signals in TIC quantification will be mitigated. Furthermore, by decreasing interference from misaligned voxels, the fitted curve to the time-intensity curve should result in a better quality of fit, as measured by $R^2$ and the sum of squared errors.

For the time-intensity curve analysis, all of the expected observations were verified following generation of bolus time-intensity curves and quantitative metrics. Following MC for every patient's imaging study, the SSE decreased ($p = 0.065$), while the RMSE error decreased significantly ($p = 0.018$). The $R^2$ value for the fitted curve increased significantly ($p = 0.029$). These trends highlight how the MC algorithm improves TIC curve fitting by decreasing the variance in the intensity signals of the imaging acquisition that are used for perfusion quantification. Additionally, for some imaging acquisitions, the maximum intensity observed pre-MC reaches values nearly three-times greater than the maximum intensity observed post-MC, as seen in **Figure 6**. This indicates that the MC algorithm is able to improve the reliability of quantification findings by mitigating the number of occurrences where non-lesion intensity values are interpreted as tumor perfusion, vasculature, and angiogenesis.

The advent of DCE-US imaging in the context of clinically monitoring tumor angiogenesis and microvasculature and evaluating response to anti-angiogenic treatment has been fueled by promising results observed in many preclinical and clinical studies investigating the utility of the imaging modality. Preclinical studies have been able to demonstrate that the vascular pathology and tumor microvasculature perfusion determined by CEUS quantification is concordant with the actual histopathology of the tumor (22), while clinical studies have successfully differentiated between malignant and benign tumors, as well as responding and non-responding tumors to anti-angiogenic therapy, by employing TIC analysis to measure kinetic properties of tumor perfusion and identify morphological differences between tumors (23-24). Having verified the qualities of our MC algorithm in clinical data sets, future work will aim to test our algorithm directly in the clinical setting and determine whether it has a translational effect in improving the assessment of tumor response to treatment. Parametric mapping and time-intensity curve analysis have been able to successfully characterize to a limited degree, so it is our hope that the MC algorithm's ability to significantly decrease the poor signal-to-noise ratio within the lesion's ROI, improve the similarity of images on a frame-to-frame basis, and elucidate the tumor's morphology will translate into more accurate and confident prognostications of patient progress during cancer treatment.

## VI. Conclusion

Investigations into the efficacy of 3D DCE-US imaging for cancer treatment monitoring have shown that 3D DCE-US has the potential to become an effective imaging modality for cancer treatment prognosis (1) (4) (5). However, the presence of motion artifacts on imaging studies severely inhibits the quality of quantification, thus making MC of 3D DCE-US imaging a crucial target for further investigation. This study presents and validates a MC algorithm with great potential for mitigating motion artifacts in 3D DCE-US. We demonstrate that the algorithm consistently generates improvement across a variety of different liver metastasis imaging acquisitions qualitatively, and highlight its robustness quantitatively through improvement in lesion volume overlap, increase in image similarity, and better accuracy of bolus-time intensity analysis. These results indicate that the utilization of the MC algorithm can help mitigate motion artifacts, thereby improving quantification of 3D DCE-US features and evaluation of tumor response to treatment and allowing physicians to develop personalized therapies based off treatment response with greater potential for recovery.